\documentclass[11pt,a4paper]{article}
\usepackage[hyperref]{emnlp2018}
\usepackage{times}
\usepackage{latexsym}

\usepackage[colorinlistoftodos]{todonotes}
\usepackage{CJKutf8}
\usepackage{comment}
\usepackage{amssymb}
\usepackage{booktabs}
\usepackage{multirow}
\usepackage{soul}
\usepackage{graphicx}
\usepackage{lscape}
\usepackage{subcaption}
\usepackage{epsfig}
\usepackage{amsmath}
\usepackage{booktabs}
\usepackage{color}
\usepackage{enumitem}
\usepackage{ifsym}
\usepackage{marvosym}
\usepackage{float}
\usepackage{url}

\newcommand{\tf}{\ensuremath{\mathit{tf}}}

\usepackage{url}
\usepackage{tabularx}
\aclfinalcopy

\title{NPRF: A Neural Pseudo Relevance Feedback Framework for Ad-hoc Information Retrieval}

\author{Canjia Li$^1$, Yingfei Sun$^1$, Ben He$^{1,3}$\textsuperscript{\Letter}, Le Wang$^{1,4}$, Kai Hui$^2$,\\ 
	\textbf{Andrew Yates$^5$, Le Sun$^{3}$, Jungang Xu$^1$} \\
	$^{1}$ University of Chinese Academy of Sciences, Beijing, China \hspace{0.1cm} $^{2}$ SAP SE, Berlin, Germany\\ 
	$^{3}$ Institute of Software, Chinese Academy of Sciences, Beijing, China\\
    $^{4}$ Computer Network Information Center, Chinese Academy of Sciences, Beijing, China\\
    $^5$ Max Planck Institute for Informatics, Saarbr\"ucken, Germany \\
\small {\tt \{licanjia17, wangle315\}@mails.ucas.ac.cn}, \hspace{0.15cm}  {\tt \{yfsun, benhe\textsuperscript{\Letter}, xujg\}@ucas.ac.cn}  \\ 
\small  {\tt kai.hui@sap.com} , \hspace{0.15cm} {\tt ayates@mpi-inf.mpg.de}, \hspace{0.15cm} {\tt sunle@iscas.ac.cn}
}
\begin{document}
\maketitle
\begin{abstract}
Pseudo relevance feedback (PRF) is commonly used to boost the performance of traditional information retrieval (IR) models by using top-ranked documents to identify and weight new query terms, thereby reducing the effect of query-document vocabulary mismatches. While neural retrieval models have recently demonstrated strong results for ad-hoc retrieval, combining them with PRF is not straightforward due to incompatibilities between existing PRF approaches and neural architectures. To bridge this gap, we propose an end-to-end neural PRF framework that can be used with existing neural IR models by embedding different neural models as building blocks. Extensive experiments on two standard test collections confirm the effectiveness of the proposed NPRF framework in improving the performance of two state-of-the-art neural IR models.
\end{abstract}

\def \vignore {0mm}

\vspace{\vignore}
\section{Introduction}\label{sec:Introduction}
\vspace{\vignore}

Recent progress in neural information retrieval models (NIRMs) has highlighted promising performance
on the ad-hoc search task. State-of-the-art NIRMs, such as DRMM~\cite{DBLP:conf/cikm/GuoFAC16}, HiNT~\cite{DBLP:conf/sigir/FanGLXZC18},
(Conv)-KNRM~\cite{DBLP:conf/sigir/XiongDCLP17,DBLP:conf/wsdm/DaiXC018},
and (Co)-PACRR~\cite{DBLP:conf/emnlp/HuiYBM17,DBLP:conf/wsdm/HuiYBM18}, 
have successfully implemented insights from  traditional IR models using  
neural building blocks.
Meanwhile, 
existing IR research has already demonstrated the effectiveness of 
incorporating relevance signals from top-ranked documents through 
pseudo relevance feedback (PRF) models~\cite{buckley08rfoverview,DBLP:conf/acl/0001MC16}.
PRF models expand the query with terms selected from top-ranked documents, thereby
boosting ranking performance by reducing the problem of vocabulary mismatch between the original query and documents~\cite{rocchio1971relevance}.
Existing neural IR models do not have a mechanism for treating expansion terms differently from the original query terms, however, making it non-trivial to combine them with existing PRF approaches.
In addition, 
neural IR models differ in their architectures, 
making the development of 
a widely-applicable PRF approach a challenging task.

To bridge this gap, we propose a generic neural pseudo relevance feedback framework, coined NPRF,
that enables the use of PRF with existing neural IR models.
Given a query and a target document, the top-ranked documents from the initial ranking
are consumed by NPRF, which
expands the query by interpreting it from different perspectives.
Given a target document to evaluate,
NPRF produces a final relevance score by considering the target document's relevance
to these top-ranked documents and to the original query.

The proposed NPRF framework can directly incorporate different established neural IR models,
which serve as the concrete scorers in evaluating the relevance of a document relative to 
the top-ranked documents and to the query, without changing their architectures.
We instantiate the NPRF framework using two state-of-the-art neural IR models, and 
we evaluate their performance on two widely-used TREC benchmark datasets for ad-hoc retrieval. 
Our results
confirm that the NPRF framework can substantially improve the performance of both models.
Moreover, both neural models perform similarly inside the NPRF framework despite the fact
that without NPRF one model performed substantially worse than the other model.
The contributions of this work are threefold: 1) the novel NPRF framework; 2)
two instantiations of the NPRF framework using
two state-of-the-art neural IR models;
and 3) the experiments that confirm the effectiveness of the
NPRF framework.

The rest of this paper is organized as follows. Section~\ref{sec:Method} presents the proposed NPRF framework in details. Following that, Section~\ref{sec.evaluation} describes the setup of the evaluation, and reports the results. Finally, Section~\ref{sec:related} recaps existing literature, before drawing conclusions in Section~\ref{sec:Conclusions}.
\begin{figure*}
    \centering
      \includegraphics[width=0.8\textwidth,height=7cm]{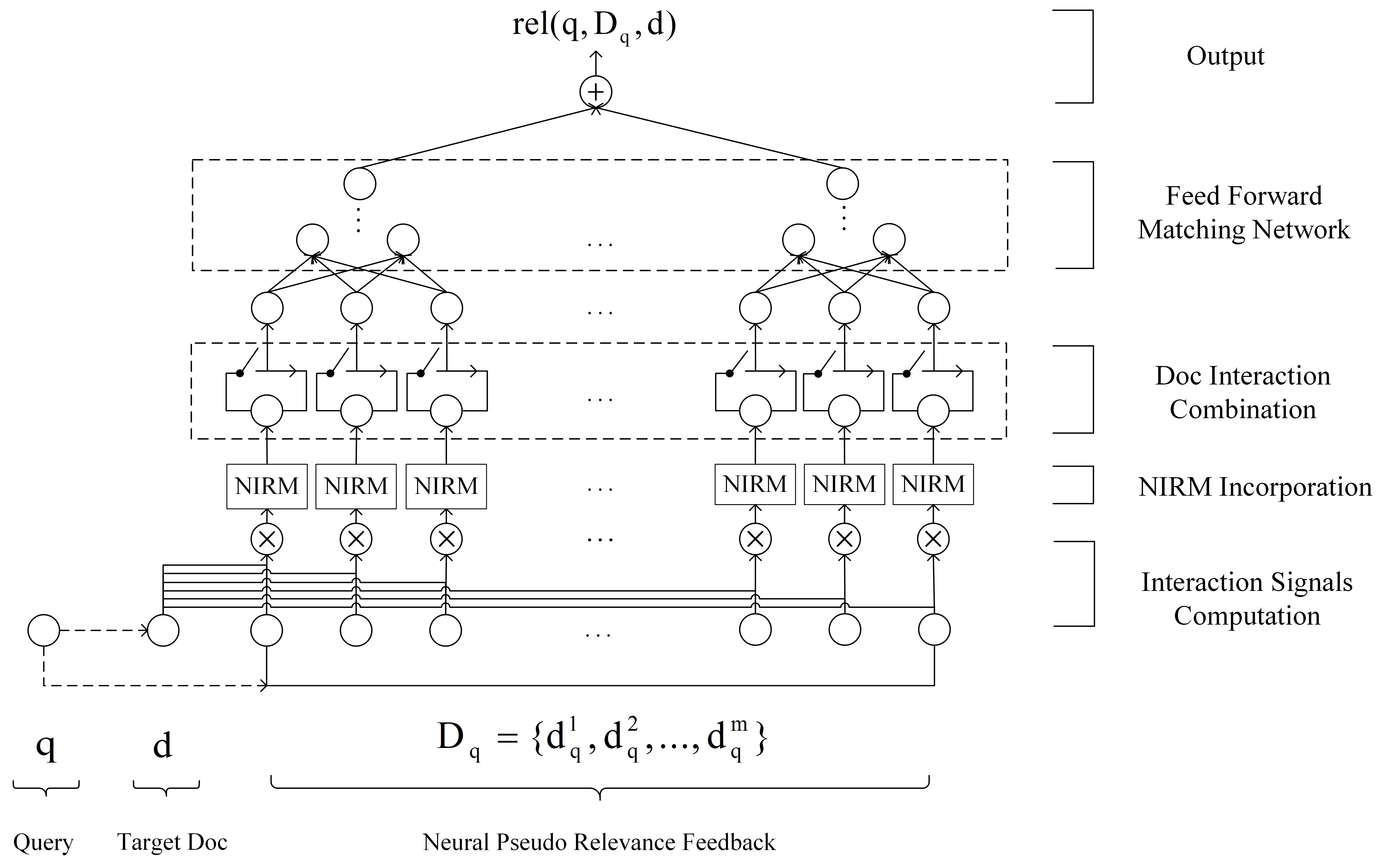}
    \caption{Architecture of the proposed neural pseudo relevance feedback (NPRF) framework.}\label{fig.cprfarch}
\end{figure*}

\vspace{\vignore}
\section{Method}\label{sec:Method}
\vspace{\vignore}
In this section, we introduce the proposed neural framework for pseudo relevance feedback (NPRF).
Recall that existing unsupervised PRF models~\cite{rocchio1971relevance,DBLP:conf/sigir/LavrenkoC01,DBLP:conf/trec/YeHHYL09} issue a query 
to obtain an initial ranking, 
identify promising terms from the top-$m$ documents returned,
and expand the original query with these terms.
Rather than selecting the expanded terms within the top-$m$ documents, NPRF uses these documents directly 
as expansion queries by considering the interactions between them and a target document. Thus, each document's ultimate relevance score depends on both its interactions with the original query
and its interactions with these feedback documents.

\vspace{\vignore}
\subsection{Overview}\label{sec.methodoverview}
\vspace{\vignore}

Given a query $q$, NPRF estimates the relevance of a target document $d$ relative to $q$ as described in the following steps. The architecture is summarized in Figure~\ref{fig.cprfarch}.
Akin to the established neural IR models like DRMM~\cite{DBLP:conf/cikm/GuoFAC16},
the description is based on a query-document pair,
and a ranking can be produced by sorting the documents according to their scores.

\begin{enumerate}[leftmargin=*]
\item[-]  
\noindent \textbf{Create initial ranking.} 
Given a document corpus,
a ranking method $\mathit{rel_q}(q, d)$ is applied to
individual documents to obtain the top-$m$ documents,
denoted as $D_q$ for $q$.

\item[-] 
\noindent \textbf{Extract document interactions.} To evaluate the 
relevance of $d$, 
each $d_q$ in $D_q$
is used to expand $q$, where $d$ is compared against
each $d_q$, using a ranking method $\mathit{rel_d}(d_q, d)$.

\item[-] 
\noindent \textbf{Combine document interactions.} The relevance scores $\mathit{rel}_{d}(d_q, d)$ for
individual $d_q\in D_q$ are further weighted by $\mathit{rel_q}(q, d_q)$, which serves as an estimator for  
the confidence of the contribution of $d_q$ relative to $q$.
The weighted combination of these relevance scores is used to produce
a relevance score for $d$, denoted as $\mathit{rel_D}(q, D_q, d)$.
\end{enumerate}

While the same ranking model can be used for both $\mathit{rel_q}(., .)$ and $\mathit{rel_d}(., .)$, we denote them separately in the architecture.
In our experiments, the widely-used unsupervised ranking method
BM25~\cite{DBLP:conf/trec/RobertsonWHGP95} serves as $\mathit{rel_q}(., .)$;
meanwhile
two state-of-the-art neural IR relevance matching models, namely,
DRMM~\cite{DBLP:conf/cikm/GuoFAC16} and K-NRM~\cite{DBLP:conf/sigir/XiongDCLP17},
serve as the ranking method $\mathit{rel_d}(., .)$.
However, it is worth noting that in principle
$\mathit{rel_q}$ and $\mathit{rel_d}$ can be replaced with any ranking method,
and the above choices mainly aim to demonstrate the effectiveness of the NPRF framework.

\vspace{\vignore}
\subsection{Model Architecture}\label{sec.architecture}
\vspace{\vignore}

The NPRF framework begins with an initial ranking for the input query $q$ determined by $\mathit{rel_q}(., .)$, which forms $D_q$, the set of the top-$m$ documents $D_q$.  
The ultimate query-document relevance score $\mathit{rel_D}(q, D_q, d)$ is computed as follows.

\textbf{Extracting document interactions.}
Given the target document $d$ and each feedback document $d_q\in D_q$,
$\mathit{rel_d}(., .)$ is used to evaluate the relevance between $d$ and $d_q$,
resulting in $m$ real-valued relevance scores, where each score corresponds to the estimated relevance of $d$ according to one feedback document $d_q$.

As mentioned, two NIRMs
are separately used to compute $\mathit{rel}_{d}(d_q, d)$ in our experiments.
Both models take as input the cosine similarities between each pair of terms in $d_q$ and $d$, which are computed using
pre-trained word embeddings as explained in Section \ref{sec.evalsetup}. Given that both models consider only unigram matches and do not consider term dependencies,
we first summarize $d_q$ by retaining only the 
top-$k$ terms according to their \tf-$idf$ scores,
which speeds up training by reducing the document size and removing noisy terms.
In our pilot experiments, the use of top-$k$ \tf-$idf$ document summarization did not influence performance.
For different $d_q\in D_q$, 
the same model is used as $\mathit{rel}_{d}(., .)$ for different pairs of $(d_q,d)$ by sharing model weights.

\textbf{Combining document interactions.} 
When determining the relevance of a target document $d$,
there exist two sources of relevance signals to consider: the target document's relevance relative to
the feedback documents $D_q$ and its relevance relative to the query $q$ itself.
In this step, we combine $\mathit{rel}_{d}(d_q, d)$ for each $d_q\in D_q$
into an overall feedback document relevance score $\mathit{rel_D}(q, D_q, d)$.
When combining the relevance scores,
the agreement between $q$ and each $d_q$ is also important, since $d_q$ may differ from $q$ in terms of information needs. 
The relevance of $d_q$ from the initial ranking $\mathit{rel}_{q}(q, d_q)$ is 
employed to quantify this agreement and weight each $\mathit{rel}_{d}(d_q, d)$ accordingly.

When computing such agreements, it is necessary to remove the influence of the absolute ranges of the scores from the initial ranker.
For example, ranking scores from a language model~\cite{DBLP:conf/sigir/PonteC98} and from BM25~\cite{DBLP:conf/trec/RobertsonWHGP95} can differ substantially in their absolute ranges.
To mitigate this, we use a smoothed min-max normalization to rescale 
$\mathit{rel}_{q}(q, d_q)$ into the range $[0.5,1]$.
The min-max normalization is applied by considering
$\mathit{min}(\mathit{rel}_{q}(q, d_q)|d_q\in D_q)$ and 
$\mathit{max}(\mathit{rel}_{q}(q, d_q)|d_q\in D_q)$.
Hereafter, 
 $\mathit{rel}_{q}(q, d_q)$ is used to denote this
relevance score after min-max normalization for brevity.
The (normalized) relevance score is smoothed and then weighted by the relevance evaluation of $d_q$,
producing a weighted document relevance score $\mathit{rel_d}^\prime(d_q, d)$ for each $d_q\in D_q$
that reflects the relevance of $d_q$ relative to $q$.
This computation is described in the following equation.

\begin{equation}\label{eq.gating}
\mathit{rel_d}^\prime(d_q, d)=\mathit{rel_d}(d_q, d)(0.5+0.5\times \mathit{rel}_{q}(q, d_q))
\end{equation}

As the last step, we propose two variants
for combining the $\mathit{rel_d}^\prime(d_q, d)$ for different $d_q$ into a single score $\mathit{rel_D}(q, D_q, d)$:
(i) performing a direct summation and (ii) using a feed forward network with a hyperbolic tangent ($\mathit{tanh}$)
non-linear activation.
Namely,
the first variant simply sums up the scores, whereas the second
takes the ranking positions of individual feedback documents into account. 

\vspace{\vignore}
\subsection{Optimization and Training}
\vspace{\vignore}
Each training sample consists of a query $q$, a set of $m$ feedback documents $D_q$, 
a relevant target document $d^+$ and a non-relevant target document $d^-$
according to the ground truth.
The Adam optimizer~\cite{adam} is used with a learning rate $0.001$ and a batch size of 20. Training normally converges within 30 epochs, with weights uniformly initialized.
A hinge loss is employed for training 
as shown below.
\begin{align}\label{eq.loss}
\begin{split}
&\mathit{loss}(q, D_q, d^+, d^-) =  \\
&max(0, 1 - \mathit{rel}(q, D_q, d^+)+\mathit{rel}(q, D_q, d^-)) \nonumber
\end{split}
\end{align}

\vspace{\vignore}
\section{Evaluation}\label{sec.evaluation}
\vspace{\vignore}
\subsection{Evaluation Setup}\label{sec.evalsetup}
\vspace{\vignore}

\noindent {\bf Dataset.} 
We evaluate our proposed NPRF framework on two standard test collections, namely, \textsc{TREC}1-3 \cite{DBLP:conf/sigir/Harman93} and Robust04~\cite{DBLP:conf/trec/Voorhees04b}. TREC1-3 consists of 741,856 documents with 150 queries used in the TREC 1-3 ad-hoc search tasks \cite{DBLP:conf/sigir/Harman93,DBLP:conf/trec/Harman94,DBLP:journals/ipm/Harman95a}. Robust04 contains 528,155 documents and 249 queries used in the TREC 2004 Robust track~\cite{DBLP:conf/trec/Voorhees04b}. We use those two collections to balance between the number of queries and the TREC pooling depth, i.e., 100 on both collections, allowing for sufficient training data. Manual relevance judgments are available on both collections, where both the relevant and non-relevant documents are labeled for each query.

Two versions of queries are included in our experiments: a short keyword query (\textit{title query}), and a longer description query that restates the corresponding keyword query's information need in terms of natural language (\textit{description query}). We evaluate each type of query separately using the metrics Mean Average Precision at 1,000 (MAP), Precision at 20 (P@20) \cite{DBLP:books/daglib/0021593}, and NDCG@20~\cite{Jarvelin:2002:CGE:582415.582418}.

\noindent \textbf{Preprocessing.} Stopword removal and Porter's stemmer are applied \cite{DBLP:books/daglib/0021593}. The word embeddings are pre-trained based on 
a pool of the top 2,000 documents returned by BM25 for individual queries
as suggested by~\cite{DBLP:conf/acl/0001MC16}.
The implementation of Word2Vec\footnote{https://code.google.com/p/word2vec/} 
from~\cite{word2vec} is employed.
In particular, we employ CBOW with the dimension set to 300, window size to 10, minimum count to 5,
and a subsampling threshold of $10^{-3}$.
The CBOW model is trained for 10 iterations on the target corpus.

\noindent\textbf{Unsupervised ranking models} serve as baselines for comparisons. We use the open source Terrier platform's~\cite{terrier12osir} implementation of these ranking models:

\begin{itemize}[leftmargin=*]
\item[-] 

\noindent {\bf BM25}~\cite{DBLP:conf/trec/RobertsonWHGP95}, a classical probabilistic model, is employed as an unsupervised baseline. The hyper-parameters {\em b} and {\em $k_1$} are tuned by grid search. As mentioned in Sec.~\ref{sec.methodoverview}, BM25 also generates the initial rankings $D_q$, serving as $rel_q(.,.)$ in the NPRF framework.

\item[-] 
\noindent On top of BM25, we use an adapted version of Rocchio's query expansion~\cite{DBLP:conf/trec/YeHHYL09}, denoted as {\bf BM25+QE}. 
Note that, as demonstrated in the results, BM25+QE's performance is comparable with the base neural IR models, including DRMM, K-NRM and PACRR. This illustrates
the difficulty in making improvements on the TREC benchmarks
through the uses of deep learning methods.
The hyper-parameters, including
the number of feedback documents and the number of expansion terms, are optimized using grid search on training queries.

\item[-] 
\noindent In addition, {\bf QL+RM3}, the query likelihood language model with the popular RM3 PRF \cite{DBLP:conf/sigir/LavrenkoC01}, is used as another unsupervised baseline.

\end{itemize}

\noindent\textbf{Neural IR models} are used for $rel_d(.,.)$. As mentioned in Section~\ref{sec.methodoverview},
two unigram neural IR models are employed in our experiments:

\begin{itemize}[leftmargin=*]

\item[-] 
\noindent \textbf{DRMM}. We employ the variant with the best effectiveness on Robust04
according to \cite{DBLP:conf/cikm/GuoFAC16}, namely, 
DRMM$_{LCH\times IDF}$ with the original configuration.

\item[-] 
\noindent \textbf{K-NRM}.
Due to the lack of training data compared with the commercial data used by~\cite{DBLP:conf/sigir/XiongDCLP17}, we employ a K-NRM variant with a frozen word embedding layer.
To compensate for this substantial reduction in the number of learnable weights,
we add an additional fully connected layer to the model.
These changes lead to a small but competitive K-NRM variant, as demonstrated in~\cite{DBLP:conf/wsdm/HuiYBM18}.

\item[-]
\noindent We additionally implement \textbf{PACRR}~\cite{DBLP:conf/emnlp/HuiYBM17}
for the purpose of performing comparisons, but do not use PACRR to compute
$rel_d(.,.)$ due to the computational costs.
In particular, PACRR-firstk is employed where the first $1,000$ terms
are used to compute the similarity matrices, and the original configuration from~\cite{DBLP:conf/emnlp/HuiYBM17} is used.

\item[-] 
\noindent \textbf{NIRM(QE)} uses the modified query generated by the query expansion of BM25+QE \cite{DBLP:conf/trec/YeHHYL09} 
as input to the neural IR model. Both DRMM and K-NRM are used to instantiate NIRM(QE).

\item[-]
\noindent \textbf{Variants of the proposed NPRF approach.}
As indicated in Section~\ref{sec.architecture},
NPRF includes two variants that differ in the combination of the relevance scores from different $d_q\in D_q$:
the variant {\bf NPRF}$\mathbf{_\ensuremath{\mathbf{ff}}}$ 
uses a feed forward network with a hidden layer with five neurons to compute $rel(d, D_q)$,
and the other variant {\bf NPRF}$\mathbf{_{ds}}$ performs a direct summation of the different relevance scores.
For the purposes of comparison, we additionally
introduce another variant coined {\bf NPRF}$\mathbf{_{\ensuremath{\mathbf{ff}}^\prime}}$, where 
the relevance of $d_q$ to $q$ is not considered in the combination by 
directly setting $rel_d^\prime(d, d_q)=rel_d(d, d_q)$ in place of
Equation~\ref{eq.gating}, 
thereafter combining the scores with a fully connected layer as in  NPRF$_{\ensuremath{\mathit{ff}}}$.
We combine each of the three NPRF variants with the DRMM and K-NRM models,
and report results for all six variants.
Our implementation of the NPRF framework is available to enable future comparisons\footnote{\url{https://github.com/ucasir/NPRF}}.

\end{itemize}

Akin to \cite{DBLP:conf/cikm/GuoFAC16,DBLP:conf/sigir/XiongDCLP17,DBLP:conf/emnlp/HuiYBM17}, the NIRM baselines and the proposed NPRF are employed to re-rank the search results from BM25.
In particular,
the top-$10$ documents from the unsupervised baseline are used as the pseudo relevance feedback documents $D_q$ as input for NPRF,
where each $d_q \in D_q$ is represented by its top-$20$ terms with the highest \tf-$idf$ weights.
As illustrated later in Section \ref{sec.analysis}, NPRF's performance is stable over a wide range of settings for both parameters.

\noindent {\bf Cross-validation.} 
Akin to~\cite{DBLP:conf/wsdm/HuiYBM18},
owing to the limited number of 
labeled data,
five-fold cross-validation is used to report the results by randomly splitting all queries into five equal partitions.
In each fold,
three partitions are used for training, one for validation, and one for testing. The model with the best MAP on the validation set is selected.
We report the average performance on all \textit{test} partitions. A two-tailed paired t-test 
is used to report the statistical significance at 95\% confidence interval.

\begin{table*}[tbh]
\centering
\resizebox{\textwidth}{!}{
\begin{tabular}{l|lr|lr|lr|lr|lr|lr}
\hline \hline
 & \multicolumn{6}{c}{Title} & \multicolumn{6}{c}{Description} \\
\cline{2-13}
Model  & \multicolumn{2}{c}{MAP}    & \multicolumn{2}{c}{P@20}   & \multicolumn{2}{c}{NDCG@20} & \multicolumn{2}{c}{MAP}    & \multicolumn{2}{c}{P@20}   & \multicolumn{2}{c}{NDCG@20} \\
\hline \hline
BM25       & 0.2408 &-& 0.4803 &-&0.4947		&-	&	0.2094&-&	0.4613&-&	0.4838 &- \\
\hline
NPRF$_{\ensuremath{\mathit{ff}}}$-DRMM           & $0.2669^\dagger$ & 10.85\% &	0.5010 & 4.31\% &	0.5119 & 3.47\%	&	$0.2509^{\dagger}$ & 19.80\% &	$0.5257^{\dagger}$ & 13.95\%&	$0.5393^{\dagger}$ & 11.46\%  \\
NPRF$_{\ensuremath{\mathit{ff}}^\prime}$-DRMM    & $0.2671^\dagger$ & 10.93\% &	$0.5023^\dagger$ & 4.59\% &	0.5116 & 3.42\%	&	$0.2504^\dagger$ & 19.58\%&	$0.5163^\dagger$ & 11.93\%&	$0.5291^\dagger$ & 9.37\%  \\
NPRF$_{ds}$-DRMM           & $0.2698^\dagger$ & 12.03\% &	$0.5187^\dagger$ & 7.99\% &	$0.5282^\dagger$ & 6.77\%	&	$\textbf{0.2527}^\dagger$ & 20.67\%&	$\textbf{0.5283}^\dagger$ & 14.53\%&	 $0.5444^\dagger$ & 12.52\%  \\
\hline
NPRF$_{\ensuremath{\mathit{ff}}}$-KNRM       & $0.2633^\dagger$ & 9.34\%   & $0.5033$ & 4.80\% & $0.5171$ & 4.52\%	&	$0.2486^{\dagger}$ & 18.71\% & 	$0.5240^{\dagger}$ & 13.59\% & $0.5398^{\dagger}$ & 11.58\% \\
NPRF$_{\ensuremath{\mathit{ff}}^\prime}$-KNRM    & $0.2654^{\dagger}$ & 10.22\%&	$0.5077^{\dagger}$ & 5.70\% &	$0.5216^{\dagger}$ & 5.44\%	&	$0.2462^{\dagger}$ & 17.60\%&	$0.5197^{\dagger}$ & 12.65\%&	$0.5363^{\dagger}$ & 10.84\% \\
NPRF$_{ds}$-KNRM    & $\textbf{0.2707}^{\dagger}$ & 12.41\%&	$\textbf{0.5303}^{\dagger}$ & 10.42\%&	$\textbf{0.5406}^{\dagger}$ & 9.29\%&	$0.2505^{\dagger}$ & 19.61\%&	$0.5270^{\dagger}$ & 14.24\%&	$\textbf{0.5460}^{\dagger}$ & 12.87\%\\
\hline
\end{tabular}}
\caption{\small Comparisons between NPRF and \textit{BM25} on \textit{TREC1-3} dataset. Relative performances compared with \textit{BM25} are in percentages. Significant improvements relative to the baselines are marked with $\dagger$.}\label{tab.BM25_d12}
\end{table*}

\begin{table*}[tbh]
\centering
\resizebox{\textwidth}{!}{
\begin{tabular}{l|lr|lr|lr|lr|lr|lr}
\hline \hline
 & \multicolumn{6}{c}{Title} & \multicolumn{6}{c}{Description} \\
\cline{2-13}
Model  & \multicolumn{2}{c}{MAP}    & \multicolumn{2}{c}{P@20}   & \multicolumn{2}{c}{NDCG@20} & \multicolumn{2}{c}{MAP}    & \multicolumn{2}{c}{P@20}   & \multicolumn{2}{c}{NDCG@20} \\
\hline \hline
BM25       & 0.2533 &-& 0.3612 &  -& 0.4158  & - & 0.2479 &  -&0.3514 & - &0.4110   &- \\
\hline
NPRF$_{\ensuremath{\mathit{ff}}}$-DRMM    & $0.2823^\dagger$ &11.46\% & $0.3941^{\dagger}$  & 9.11\%& $0.4350^\dagger$  & 4.62\%& $0.2766^{\dagger}$ & 11.58\% & $0.3908^{\dagger}$ & 11.21\% & $0.4421^{\dagger}$   & 7.56\% \\
NPRF$_{\ensuremath{\mathit{ff}}^\prime}$-DRMM & $0.2837^{\dagger}$          & 12.00\% & $0.3928^\dagger$            & 8.74\%  & $0.4377^\dagger$            & 5.27\% & $0.2774^{\dagger}$          & 11.90\% & $0.3984^{\dagger}$          & 13.38\% & $0.4493^{\dagger}$          & 9.32\%  \\
NPRF$_{ds}$-DRMM                              & $\textbf{0.2904}^{\dagger}$ & 14.66\% & $\textbf{0.4064}^{\dagger}$ & 12.52\% & $\textbf{0.4502}^{\dagger}$ & 8.28\% & $\textbf{0.2801}^{\dagger}$ & 12.95\% & $\textbf{0.4026}^{\dagger}$ & 14.57\% & $\textbf{0.4559}^{\dagger}$ & 10.92\% \\
\hline
NPRF$_{\ensuremath{\mathit{ff}}}$-KNRM        & $0.2809^{\dagger}$          & 10.90\% & $0.3851^{\dagger}$          & 6.62\%  & $0.4287$          & 3.11\% & $0.2720^{\dagger}$          & 9.71\%  & $0.3867^{\dagger}$          & 10.06\% & $0.4356^{\dagger}$          & 5.99\%  \\
NPRF$_{\ensuremath{\mathit{ff}}^\prime}$-KNRM & $0.2815^{\dagger}$          & 11.13\% & $0.3882^{\dagger}$          & 7.48\%  & $0.4264$          & 2.55\% & $0.2737^{\dagger}$          & 10.39\% & $0.3892^{\dagger}$          & 10.74\% & $0.4382^{\dagger}$          & 6.61\%  \\
NPRF$_{ds}$-KNRM                              & $0.2846^{\dagger}$          & 12.36\% & $0.3926^{\dagger}$          & 8.69\%  & $0.4327$          & 4.06\% & $0.2800^{\dagger}$          & 12.95\% & $0.3972^{\dagger}$          & 13.03\% & $0.4477^{\dagger}$          & 8.94\% \\
\hline
\end{tabular}}
\caption{\small Comparisons between NPRF and \textit{BM25} on the \textit{Robust04} dataset. Relative performances compared with \textit{BM25} are in percentages. Significant improvements relative to the baselines are marked with $\dagger$.}\label{tab.BM25_rob}
\end{table*}

\begin{table*}[tbh]
\centering
\resizebox{\textwidth}{!}{\begin{tabular}{l|lr|lr|lr|lr|lr|lr}
\hline \hline
& \multicolumn{6}{c}{Title} & \multicolumn{6}{c}{Description} \\
\cline{2-13}
 Model  & \multicolumn{2}{c}{MAP}    & \multicolumn{2}{c}{P@20}   & \multicolumn{2}{c}{NDCG@20} & \multicolumn{2}{c}{MAP}    & \multicolumn{2}{c}{P@20}   & \multicolumn{2}{c}{NDCG@20} \\
\hline \hline
DRMM         & 0.2469&-&	0.4833&-&	0.4919&-&	0.2111&-&	0.4423&-&	0.4546&- \\
K-NRM         & 0.2284&-&	0.4410&-&	0.4530&-&	0.1763&-&	0.3753&-&	0.3854&- \\
PACRR-firstk & 0.2393&-&	0.4620&-&	0.4782&-&	0.1702&-&	0.3577&-&	0.3666&-  \\ 
\hline
NPRF$_\ensuremath{\mathit{ff}}$-DRMM        & $0.2669^{\dagger}$          & 8.12\%  & 0.5010                       & 3.66\%  & 0.5119                      & 4.06\%  & $0.2509^{\dagger}$          & 18.83\% & $0.5257^{\dagger}$          & 18.84\% & $0.5393^{\dagger}$          & 18.63\% \\
NPRF$_{\ensuremath{\mathit{ff}}^\prime}$-DRMM & $0.2671^{\dagger}$          & 8.19\%  & $0.5023$                    & 3.94\%  & 0.5116                      & 4.01\%  & $0.2504^{\dagger}$          & 18.61\% & $0.5163^{\dagger}$          & 16.73\% & $0.5291^{\dagger}$          & 16.40\% \\
NPRF$_{ds}$-DRMM        & $0.2698^{\dagger}$          & 9.26\%  & $0.5187^{\dagger}$          & 7.32\%  & $0.5282^{\dagger}$          & 7.38\%  & $\textbf{0.2527}^{\dagger}$ & 19.69\% & $\textbf{0.5283}^{\dagger}$ & 19.44\% & $0.5444^{\dagger}$          & 19.76\% \\
\hline
NPRF$_\ensuremath{\mathit{ff}}$-KNRM        & $0.2633^{\dagger}$          & 15.28\% & $0.5033^\dagger$            & 14.13\% & $0.5171^\dagger$            & 14.14\% & $0.2486^{\dagger}$          & 40.97\% & $0.5240^{\dagger}$          & 39.61\% & $0.5398^{\dagger}$          & 40.06\% \\
NPRF$_{\ensuremath{\mathit{ff}}^\prime}$-KNRM & $0.2654^{\dagger}$          & 16.20\% & $0.5077^{\dagger}$          & 15.12\% & $0.5216^{\dagger}$          & 15.15\% & $0.2462^{\dagger}$          & 39.65\% & $0.5197^{\dagger}$          & 38.45\% & $0.5363^{\dagger}$          & 39.13\% \\
NPRF$_{ds}$-KNRM        & $\textbf{0.2707}^{\dagger}$ & 18.51\% & $\textbf{0.5303}^{\dagger}$ & 20.26\% & $\textbf{0.5406}^{\dagger}$ & 19.35\% & $0.2505^{\dagger}$          & 42.04\% & $0.5270^{\dagger}$          & 40.41\% & $\textbf{0.5460}^{\dagger}$ & 41.67\% \\
\hline 
\end{tabular}}
\caption{\small Comparisons between NPRF and neural IR models on \textit{TREC1-3}. Relative performances of NPRF-DRMM(KNRM) compared with DRMM (K-NRM) are in percentages, and statistically significant improvements are marked with $\dagger$.}\label{tab.NIRM_d12}
\end{table*}

\begin{table*}[tbh]
\centering
\resizebox{\textwidth}{!}{\begin{tabular}{l|lr|lr|lr|lr|lr|lr}
\hline \hline
& \multicolumn{6}{c}{Title} & \multicolumn{6}{c}{Description} \\
\cline{2-13}
 Model  & \multicolumn{2}{c}{MAP}    & \multicolumn{2}{c}{P@20}   & \multicolumn{2}{c}{NDCG@20} & \multicolumn{2}{c}{MAP}    & \multicolumn{2}{c}{P@20}   & \multicolumn{2}{c}{NDCG@20} \\
\hline \hline
DRMM         & 0.2688 &-& 0.3713 &-& 0.4297  &-& 0.2630 &-& 0.3558 &-& 0.4135 &-\\
K-NRM         & 0.2464 &-& 0.3510 &-& 0.3989  &-& 0.1687 &-& 0.2301 &-& 0.2641 &-\\
PACRR-firstk & 0.2540 &-& 0.3631 &-& 0.4082  &-& 0.2087 &-& 0.2962 &-& 0.3362 &- \\ \hline
NPRF$_\ensuremath{\mathit{ff}}$-DRMM        & $0.2823$                    & 5.03\%  & $0.3941^{\dagger}$          & 6.14\%  & $0.4350$           & 1.24\% & $0.2766^{\dagger}$          & 5.17\%  & $0.3908^{\dagger}$          & 9.84\%  & $0.4421^{\dagger}$          & 6.92\%  \\
NPRF$_{\ensuremath{\mathit{ff}}^\prime}$-DRMM & $0.2837^{\dagger}$          & 5.55\%  & $0.3928$                    & 5.78\%  & $0.4377$           & 1.87\% & $0.2774^{\dagger}$          & 5.48\%  & $0.3984^{\dagger}$          & 11.97\% & $0.4493^{\dagger}$          & 8.67\%  \\
NPRF$_{ds}$-DRMM        & $\textbf{0.2904}^{\dagger}$ & 8.05\%  & $\textbf{0.4064}^{\dagger}$ & 9.46\%  & $\textbf{0.4502}$  & 4.78\% & $\textbf{0.2801}^{\dagger}$ & 6.46\%  & $\textbf{0.4026}^{\dagger}$ & 13.15\% & $\textbf{0.4559}^{\dagger}$ & 10.26\% \\
\hline
NPRF$_\ensuremath{\mathit{ff}}$-KNRM        & $0.2809^{\dagger}$          & 14.00\% & $0.3851^{\dagger}$          & 9.72\%  & $0.4287^{\dagger}$ & 7.48\% & $0.2720^{\dagger}$          & 61.22\% & $0.3867^{\dagger}$          & 68.08\% & $0.4356^{\dagger}$          & 64.96\% \\
NPRF$_{\ensuremath{\mathit{ff}}^\prime}$-KNRM & $0.2815^{\dagger}$          & 14.25\% & $0.3882^{\dagger}$          & 10.60\% & $0.4264^{\dagger}$ & 6.90\% & $0.2737^{\dagger}$          & 62.21\% & $0.3892^{\dagger}$          & 69.12\% & $0.4382^{\dagger}$          & 65.93\% \\
NPRF$_{ds}$-KNRM        & $0.2846^{\dagger}$          & 15.50\% & $0.3926^{\dagger}$          & 11.85\% & $0.4327^{\dagger}$ & 8.47\% & $0.2800^{\dagger}$          & 65.98\% & $0.3972^{\dagger}$          & 72.62\% & $0.4477^{\dagger}$          & 69.55\% \\
\hline
\end{tabular}}
\caption{\small Comparisons between NPRF and neural IR models on \textit{Robust04}.  Relative performances of NPRF-DRMM(KNRM) compared with DRMM (K-NRM) are in percentages, and statistically significant improvements are marked with $\dagger$.}\label{tab.NIRM_rob}
\end{table*}

\begin{table*}[tbh]
\centering
\resizebox{\textwidth}{!}{
\begin{tabular}{l|lll|lll|| lll|lll}
\hline \hline
& \multicolumn{6}{c}{TREC1-3} & \multicolumn{6}{c}{Robust04} \\
\hline
 & \multicolumn{3}{c}{Title} & \multicolumn{3}{c}{Description} & \multicolumn{3}{c}{Title} & \multicolumn{3}{c}{Description}\\
\cline{2-13}
Model  & MAP    & P@20   & NDCG@20 & MAP    & P@20   & NDCG@20 & MAP    & P@20   & NDCG@20 & MAP    & P@20   & NDCG@20\\
\hline \hline
BM25+QE & {\bf 0.2873}	& 0.5200  &	0.5330 & {\bf 0.2601}	& 0.4973	& 0.5093 & {\bf 0.2966} & 0.3839 & 0.4353 & {\bf 0.2926} & 0.3817 & 0.4340\\
QL+RM3 & 0.2734 & 0.5093 & 0.5198 & 0.2421 & 0.4627 & 0.4801 & 0.2842 & 0.3878 & 0.4398 & 0.2686 & 0.3506 & 0.4150 \\ 
DRMM (QE) & 0.2741 & 0.5183 & 0.5345 & 0.2380 & 0.5077 & 0.5229 & 0.2876 & 0.4002 & {\bf 0.4549} & 0.2711 & 0.3822 & 0.4392\\
K-NRM (QE) & 0.2633 & 0.5127 & 0.5235 & 0.2307 & 0.4877 & 0.5039 & 0.2521 & 0.3644 & 0.4062 & 0.2380  & 0.3304 & 0.3785\\ 
\hline
NPRF$_{ds}$-DRMM  & 0.2698 & 0.5187 & 0.5282 & 0.2527 & {\bf 0.5283} & 0.5444$^{\dagger}$ & 0.2904 & {\bf 0.4064} & 0.4502 & 0.2801 & {\bf 0.4026}$^{\dagger}$ & {\bf 0.4559}$^{\dagger}$\\
NPRF$_{ds}$-KNRM & 0.2707 & {\bf 0.5303} & {\bf 0.5406} & 0.2505 & 0.5270 & {\bf 0.5460}$^{\dagger}$ & 0.2846 & 0.3926 & 0.4327 & 0.2800 & $0.3972^{\dagger}$ & 0.4477\\
\hline
\end{tabular}}
\caption{\small Comparisons between NPRF and \textit{query expansion baselines} on \textit{TREC1-3} and \textit{Robust04}. Significant improvements over the best baseline is marked with $\dagger$. }\label{tab.QE}
\end{table*}
\vspace{\vignore}
\subsection{Results}\label{sec.results}
\vspace{\vignore}

\noindent {\bf Comparison to BM25}. We first compare the proposed NPRF models with the unsupervised BM25.
The results are summarized in Tables~\ref{tab.BM25_d12} and~\ref{tab.BM25_rob},
where the best result in each column is highlighted in bold. 
From Tables~\ref{tab.BM25_d12} and~\ref{tab.BM25_rob}, 
it can be seen that the proposed NPRF variants obtain significant improvement relative to BM25 
on both test collections with  both kinds of test queries. 
Moreover, the results imply that the use of different query types does not affect the effectiveness of NPRF, 
which consistently outperforms BM25.

\noindent {\bf Comparison to neural IR models}. 
NPRF is further compared with different neural IR models, as summarized in Tables~\ref{tab.NIRM_d12} \&~\ref{tab.NIRM_rob}. 
It can be seen that NPRF regularly improves on top of the NIRM baselines. 
For both types of queries, NPRF-DRMM outperforms DRMM and NPRF-KNRM outperforms K-NRM when re-ranking BM25.
Remarkably, the proposed NPRF is able to improve the weaker NIRM baseline. 
For instance, on Robust04,
when using the description queries, 
DRMM and K-NRM obtain highly different results, 
with MAPs of $0.2630$ and $0.1687$ after re-ranking the initial results from BM25, respectively.
When NPRF is used in conjunction with the NIRM models, however, 
the gap between the two models is closed; that is, 
MAP=$0.2801$ for NRFF$_{ds}$-DRMM and MAP=$0.2800$ for NRFF$_{ds}$-KNRM (see Table~\ref{tab.NIRM_rob}). 
This finding highlights that our proposed NPRF is robust with respect to the use of the two embedded NIRM models. A possible explanation 
for the poor performance of K-NRM on two TREC collections is the lack of training data, as suggested in~\cite{DBLP:conf/wsdm/DaiXC018}. While K-NRM could be improved by introducing weak supervision~\cite{DBLP:conf/wsdm/DaiXC018}, we achieve the same goal by incorporating pseudo relevance feedback information without extra training data.

While the six NPRF variants exhibit similar results across both kinds of queries, NPRF$_{ds}$-DRMM in general achieves 
the best performance on Robust04, and NPRF$_{ds}$-KNRM appears to be the best variant on TREC1-3. In the meantime,
NPRF$_{ds}$ outperforms NPRF$_\ensuremath{\mathit{ff}}$ variants. 
One difference between the two methods is that NPRF$_\ensuremath{\mathit{ff}}$ considers the position of each $d_q$ in the $D_q$ ranked documents, whereas NPRF$_{ds}$ simply sums up the scores regardless of the positions. The fact that NPRF$_{ds}$ performs better suggests
that the ranking position within the $D_q$ documents may not be a useful signal.
In the remainder of this paper, we mainly report on the results obtained by NPRF$_{ds}$.

\noindent {\bf Comparison to query expansion baselines}. In Table~\ref{tab.QE}, the proposed NPRF model is compared with three kinds of query expansion baselines, namely,
the unsupervised BM25+QE~\cite{DBLP:conf/trec/YeHHYL09}, QL+RM3 \cite{DBLP:conf/sigir/LavrenkoC01}, and DRMM/K-NRM(QE), the neural IR models using expanded queries as input. 
According to Table~\ref{tab.QE}, the unsupervised BM25+QE baseline appears to achieve
better performance in terms of MAP@1k, owing to its use of 
query expansion to match relevant documents containing the expansion terms from the whole collection. 
On the other hand, NPRF$_{ds}$, which reranks the top-1000 documents returned by BM25, outperforms the query expansion baselines in terms of early precision, as measured by either NDCG@20 or P@20. These measures on shallow rankings are particularly important for general IR applications 
where the quality of the top-ranked results is crucial to the user satisfaction. 
Moreover, our NPRF outperforms NIRM(QE) in most cases, indicating the benefit brought by wrapping up the feedback information in a document-to-document matching framework as in NPRF, as opposed to directly adding unweighted expansion terms to the query. 
Recall that, it is not straightforward to incorporate these expanded terms within the existing NIRMs' architectures because the NIRMs do not distinguish between them and the original query terms.

\vspace{\vignore}
\subsection{Analysis}\label{sec.analysis}
\vspace{\vignore}

\begin{table*}[tbh]
	\centering\small
	\begin{tabularx}{\textwidth}{|p{7.5cm}|X|}
		\hline
		\multicolumn{2}{|l|}{TREC Query 341: airport security}\\ \hline
		Terms in doc at rank {\em i} & Terms in target document FBIS3-23332 \\
		\hline
		1. terrorist detect passenger check \textcolor{blue}{police} \textcolor{blue}{scan};
		2. heathrow  terrorist \textcolor{blue}{armed} \textcolor{blue}{aviation} \textcolor{blue}{police};
		3. detect airline passenger \textcolor{blue}{police} \textcolor{blue}{scan} \textcolor{blue}{flight} \textcolor{blue}{weapon}; 4. \textcolor{blue}{aviation}; 5. detect baggage  passenger; 6. passenger bomb baggage  terrorist \textcolor{blue}{explosive} \textcolor{blue}{aviation} \textcolor{blue}{scan} \textcolor{blue}{flight} \textcolor{blue}{weapon}; 7. baggage airline detect passenger \textcolor{blue}{scan} \textcolor{blue}{flight}  \textcolor{blue}{weapon}; 8. baggage  airline passenger \textcolor{blue}{flight}; 9. passenger \textcolor{blue}{police} \textcolor{blue}{aviation}; 10. airline baggage \textcolor{blue}{aviation} \textcolor{blue}{flight}
		& transec semtex \textcolor{red}{airline} ditma {\bf security} \textcolor{red}{baggage} \textcolor{red}{heathrow} test device lockerbie klm \textcolor{red}{bomb} virgin {\bf airport} loaded blobby transport \textcolor{red}{detect} inspector \textcolor{red}{terrorist} identify atlantic depressing \textcolor{red}{passenger} fail aircraft dummy \textcolor{red}{check} inert patchy stein norwich doll regard rupert lapse busiest loophole employee campaign blew procedure traveler \textcolor{red}{passport} reconcile glasgow investigate boeing bags bag harry successive smuggle conscious reconciliation tragedy board wire hidden...\\
		\hline
	\end{tabularx}
	\caption{\small An illustrative example of soft matching in NPRF. The target document FBIS3-23332, judged relevant, is ranked 122$^{nd}$ by BM25 for query 341 on Robust04, and is promoted to the 5$^{th}$ by NPRF$_{ds}$-DRMM. The NPRF mechanism increases the chances of soft-matching query-related terms that appear in the top-ranked documents (terms in \textcolor{blue}{blue}), but are missing in both the query and the target document. Subsequently, the matching signals with the query terms (in {\bf bold}) and the query-related terms (in \textcolor{red}{red}) in the target document are enhanced.
	}\label{tab.example}
\end{table*}

\begin{figure}[!htb]
	\centering
	\begin{subfigure}[]{0.5\textwidth}
		\centering
		\includegraphics[height=2.1in]{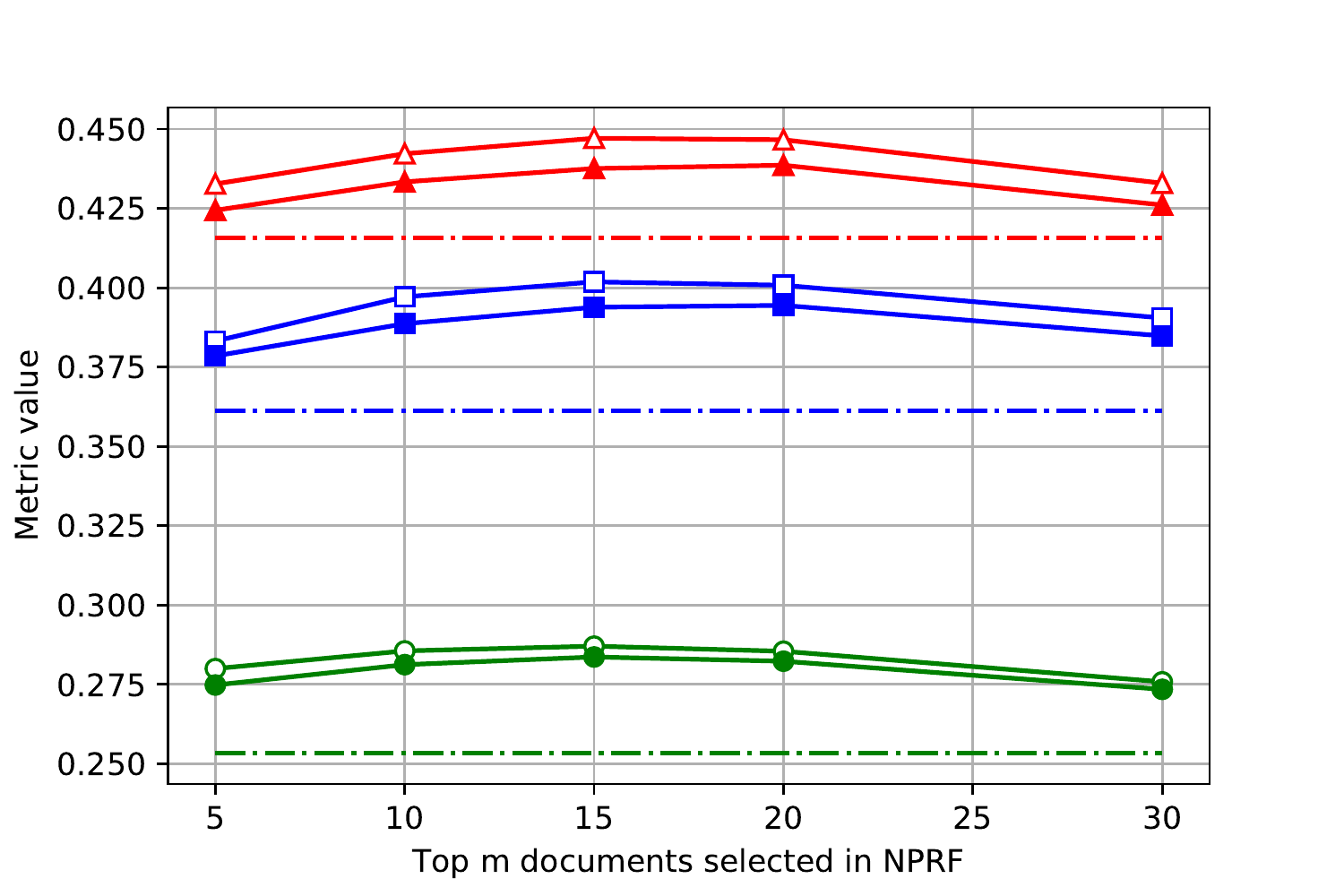}     
	\end{subfigure}%
	\\ 
	\begin{subfigure}[]{0.5\textwidth}
		\centering
		\includegraphics[height=2.1in]{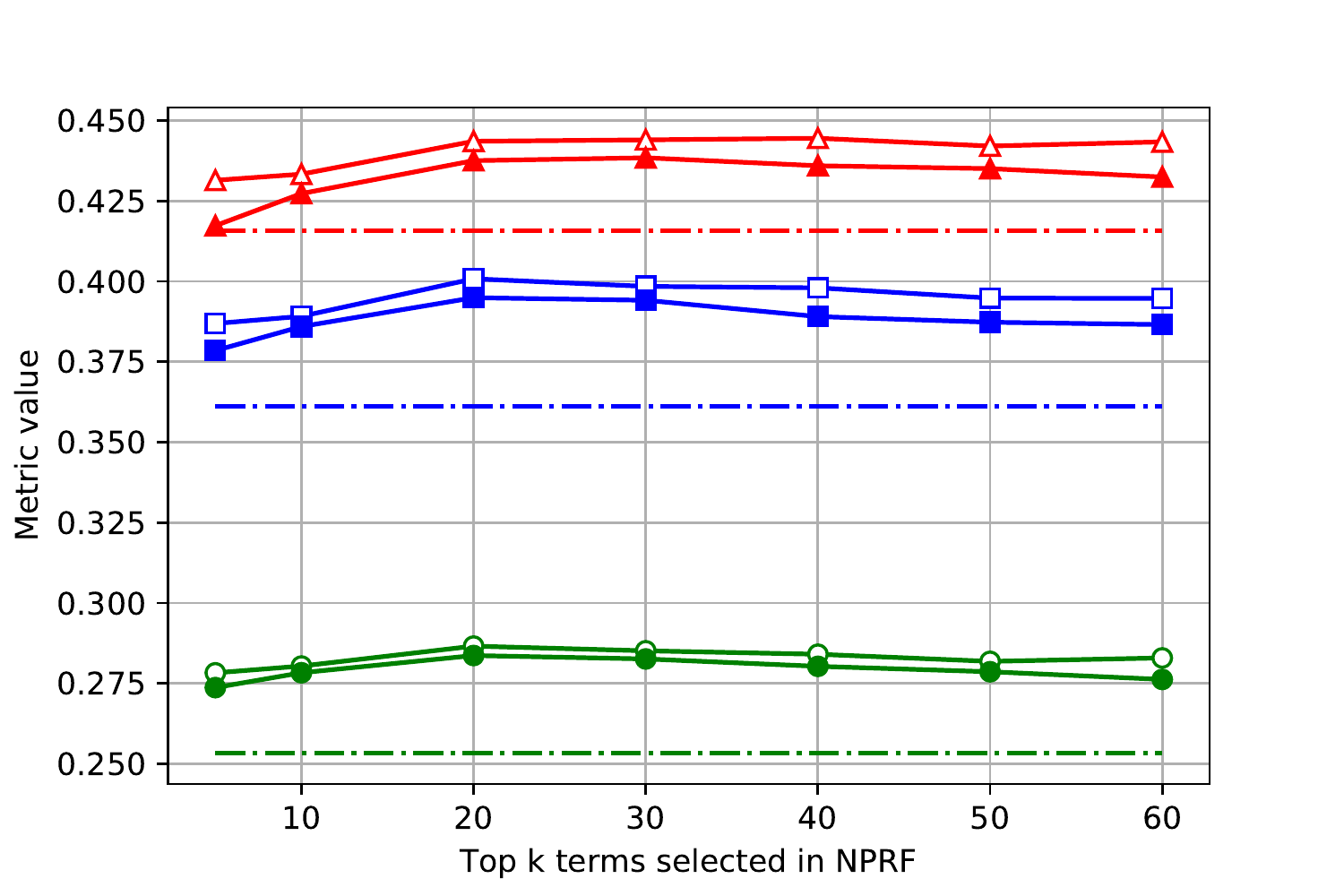}
	\end{subfigure}
	\caption{\small Performance of NPRF$_{ds}$ with different numbers of PRF documents (top) and different umber of terms which are used to summarize the feedback documents (bottom). The $\circ$, $\square$, $\bigtriangleup$ correspond to results measured by MAP, P@20 and NDCG@20 respectively, and the empty or solid symbols correspond to those for NPRF$_{ds}$-DRMM and NPRF$_{ds}$-KNRM. The three dotted lines, from bottom to top, are the BM25 baseline evaluated by MAP, P@20 and NDCG@20, respectively.}\label{fig:param}
\end{figure}

\noindent {\bf Parameter sensitivity}. Moreover, we analyze factors that may influence NPRF's performance. We report results on NPRF$_{ds}$ using title queries on Robust04 for the sake of brevity, but similar observations also hold for the other NPRF variants, as well as on TREC1-3.
Figure~\ref{fig:param} illustrates the sensitivity of NPRF relative to two parameters: the number of feedback documents $m$ within $D_q$ and the number of terms $k$ that are used to summarize each $d_q \in D_q$. Specifically, Figure~\ref{fig:param} shows the performance of NPRF$_{ds}$ as the number of feedback documents
$m$ varies (top), and as the number of top terms $k$ varies (bottom). 
The effectiveness of NPRF appears to be stable over a wide range of the parameter configurations, where the proposed model consistently outperforms the BM25 baseline.

\noindent {\bf Case study}. A major advantage of the proposed NPRF over existing neural IR models is that it allows for soft-matching query-related terms that are missing from both the query and the target document. Table~\ref{tab.example} presents an illustrative example of soft matching in NPRF. From Table~\ref{tab.example}, it can be seen that there exist query-related terms in the top-$10$ documents returned by BM25 in the initial ranking. However, since those query-related terms are missing in both the query and the target document, they are not considered in the document-to-query matching and, consequently, the target document is ranked 122$^{nd}$ by BM25 despite the facts that it was
judged relevant by a human assessor.
In contrast, the NPRF framework allows for the soft-matching of terms that are missing in both the query and target document.
As a result, the matching signals for the query terms and query-related terms in the target document are enhanced. This leads to enhanced effectiveness with the target document now ranked in the 5$^{th}$ position.

In summary, the evaluation on two standard TREC test collections shows promising results obtained by our proposed NPRF approach, which outperforms state-of-the-art neural IR models in most cases. Overall, NPRF provides effective retrieval performance that is robust with respect to the two embedded neural models used for encoding the document-to-document interactions, the two kinds of queries with varied length, and wide range of parameter configurations.

\vspace{\vignore}
\section{Related Work}\label{sec:related}
\vspace{\vignore}

Recently, several neural IR models (NIRMs) have been proposed to apply deep learning techniques in ad-hoc information retrieval. One of the essential ideas from prior work is to model the document-to-query {\em interaction} via neural networks,
based on a matrix of document-to-query embedding term similarities, 
incorporating both the ``exact matching'' of terms appearing in both the document and query and the ``soft matching'' of different query and document term pairs that are semantically related.

DSSM, one of the earliest NIRMs proposed in \cite{DBLP:conf/cikm/HuangHGDAH13}, employs a multi-layer neural network to project queries and document into a common semantic space. The cosine similarity between a query and a document (document title) is used to produce a final relevance score for the query-document pair. CDSSM is a convolutional version of DSSM, which uses the convolutional neural network (CNN) and max-pooling strategy to extract semantic matching features at the sentence level \cite{DBLP:conf/www/ShenHGDM14}. \cite{DBLP:journals/corr/PangLGXC16} also employ a CNN to construct the MatchPyramid model, which learns hierarchical matching patterns between local interactions of document-query pair. \cite{DBLP:conf/cikm/GuoFAC16} argue that both DSSM and CDSSM are representation-focused models, and thus are better suited to capturing semantic matching than relevance matching (i.e., lexical matching), and propose the interaction-focused relevance model named DRMM. DRMM maps the local interactions between a query-document pair into a fixed-length histogram, from which the exact matching signals are distinguished from the other matching signals. These signals are fed into a feed forward network and a term gating network to produce global relevance scores. Similar to DRMM, K-NRM \cite{DBLP:conf/sigir/XiongDCLP17} builds its model on top of a matrix of local interaction signals, and utilizes multiple Gaussian kernels to obtain multi-level exact/soft matching features that are input into a ranking layer to produce the final ranking score. K-NRM is later improved by Conv-KNRM, which employs CNN filters to capture $n$-gram representations of queries and documents \cite{DBLP:conf/wsdm/DaiXC018}. DeepRank \cite{DBLP:conf/cikm/PangLGXXC17} models the relevance generation process by identifying query-centric contexts, processing them with a CNN or LSTM, and aggregating them to produce a final relevance score. Building upon DeepRank, \cite{DBLP:conf/sigir/FanGLXZC18} propose to model diverse relevance patterns by a data-driven method to allow relevance signals at different granularities to compete with each other for the final relevance assessment.

Duet \cite{DBLP:conf/www/Mitra0C17} employs two separate deep neural networks to build a relevance ranking model, in which a local model estimates the relevance score according to exact matches between query and document terms, and a distributed model estimates relevance by learning dense lower-dimensional representations of query and document text. \cite{DBLP:conf/wsdm/ZamaniMSCT18} extends the Duet model by considering different fields within a document. 

\cite{DBLP:conf/emnlp/HuiYBM17} propose the PACRR model based on the idea that an appropriate combination of convolutional kernels and pooling operations can be used to successfully identify both unigram and n-gram query matches. PACRR is later improved upon by Co-PACRR, a context-aware variant that takes the local and global context of matching signals into account through the use of three new components \cite{DBLP:conf/wsdm/HuiYBM18}.

\cite{DBLP:conf/bibm/RanHHXS17} propose a document-based neural relevance model that utilizes complemented medical records to address the mismatch problem in  clinical decision support. \cite{DBLP:conf/emnlp/NogueiraC17} propose a reinforcement learning approach to reformulating a task-specific query. \cite{DBLP:conf/acl/JiCLZD18} propose DAZER, a CNN-based neural model upon interactions between seed words and words in a document for zero-shot document filtering with adversarial learning. \cite{DBLP:conf/sigir/AiBGC18} propose to refine document ranking by learning a deep listwise context model. 

In summary, most existing neural IR models are based on query-document interaction signals and do not provide a mechanism for 
incorporating relevance feedback information. This work proposes an approach for incorporating relevance feedback information by embedding neural IR models within a neural pseudo relevance feedback framework, where the models consume feedback information via document-to-document interactions.

\vspace{\vignore}
\section{Conclusions}\label{sec:Conclusions}
\vspace{\vignore}
In this work we proposed a neural pseudo relevance feedback framework (NPRF) for incorporating relevance feedback information into existing neural IR models (NIRM).
The NPRF framework uses feedback documents to better estimate relevance scores by considering individual feedback documents as different interpretations of the user's
information need. On two standard TREC datasets, NPRF significantly improves the performance of two
state-of-the-art NIRMs. Furthermore, NPRF was able to improve their performance across
two kinds of query tested (namely, short queries and the verbal queries in natural language).
Finally, our analysis demonstrated the robustness of the NPRF framework over different parameter configurations.

\section*{Acknowledgments}
This work is supported in part by the National Natural Science Foundation of China (61433015/61472391), and the Beijing Natural Science Foundation under Grant No. (4162067/4142050).

\end{document}